\documentclass[11pt]{article}

\usepackage[utf8]{inputenc} 
\usepackage[T1]{fontenc}    
\usepackage{hyperref}       
\usepackage{url}            
\usepackage{booktabs}       
\usepackage{amsfonts}       
\usepackage{amsmath}        
\usepackage{nicefrac}       
\usepackage{fancyhdr}       
\usepackage{graphicx}       
\usepackage{algorithm}      
\usepackage{algorithmic}    
\usepackage{geometry}       
\usepackage{titlesec}       

\usepackage{float}          
\usepackage{placeins}
\usepackage[font=small]{caption}
\usepackage{etoolbox} 

\graphicspath{{./}{report/}{media/}}     

\fancypagestyle{firstpage}{
    \fancyhf{}
    \fancyhead[L]{%
        \includegraphics[height=1.0cm,keepaspectratio]{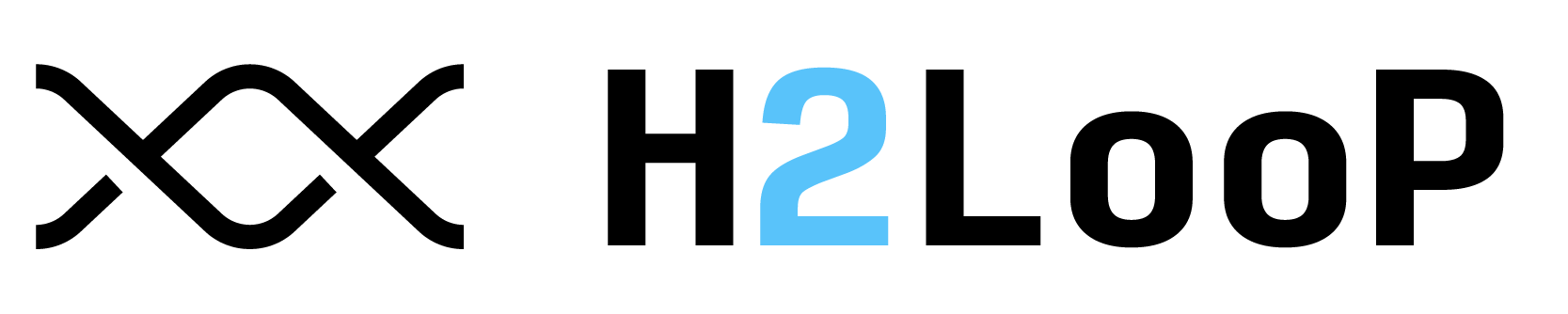}
    }
    
}

\makeatletter
\renewcommand{\maketitle}{
  \begin{center}
    {\LARGE\bfseries \@title \par}
    \vskip 1.5em
    {\large \@author \par}
  \end{center}
  \vskip 2em
}
\makeatother

\titleformat{\section}
  {\Large\bfseries}{\thesection}{1em}{}
\titleformat{\subsection}
  {\large\bfseries}{\thesubsection}{1em}{}
\titleformat{\subsubsection}
  {\normalsize\bfseries}{\thesubsubsection}{1em}{}

\title{SpecMap: Hierarchical LLM Agent for Datasheet-to-Code Traceability Link Recovery in Systems Engineering}

\author{
\begin{tabular}{c c c}
Vedant Nipane & Pulkit Agrawal & Amit Singh \\
\end{tabular}
\\[0.5em]
\textbf{H2LooP.ai}
}

\begin{document}
\thispagestyle{firstpage}

\maketitle

\begin{abstract}

Establishing precise traceability between embedded systems datasheets and their corresponding code implementations remains a fundamental challenge in systems engineering, particularly for low-level software where manual mapping between specification documents and large code repositories is infeasible. Existing Traceability Link Recovery (TLR) approaches primarily rely on lexical similarity and information retrieval techniques, which struggle to capture the semantic, structural, and symbol-level relationships prevalent in embedded systems software. We present a hierarchical datasheet-to-code mapping methodology that employs large language models (LLMs) for semantic analysis while explicitly structuring the traceability process across multiple abstraction levels. Rather than performing direct specification-to-code matching, the proposed approach progressively narrows the search space through repository-level structure inference, file-level relevance estimation, and fine-grained symbol-level alignment. The method extends beyond function-centric mapping by explicitly covering macros, structs, constants, configuration parameters, and register definitions commonly found in systems-level C/C++ codebases.
We evaluate the approach on multiple open-source embedded systems repositories using manually curated datasheet-to-code ground truth. Experimental results show substantial improvements over traditional information-retrieval-based baselines, achieving up to 73.3\% file mapping accuracy. The hierarchical decomposition significantly reduces computational overhead, lowering total LLM token consumption by 84\% and end-to-end runtime by approximately 80\%. This methodology supports automated analysis of large embedded software systems and enables downstream applications such as training data generation for systems-aware machine learning models, standards compliance verification, and large-scale specification coverage analysis.

\end{abstract}

\vskip 1em

\section{Introduction}

Modern systems engineering faces a critical challenge in establishing precise mapping between embedded systems datasheets and their code implementations. This problem is particularly acute in embedded systems, IoT devices, and standards-compliant implementations where understanding the relationship between datasheet specifications and actual code becomes essential for maintenance, verification, and development processes. The complexity of this challenge has grown significantly as embedded systems become more sophisticated and datasheet documents become more detailed.

Consider a typical embedded systems project implementing communication protocols from hardware datasheets. The datasheet specification may contain hundreds of sections describing protocols, interfaces, data structures, and behavioral requirements. The corresponding implementation spans thousands of files across complex directory hierarchies in GitHub repositories, making manual datasheet-to-code mapping virtually impossible to maintain accurately over time. This disconnect between datasheet specification and implementation creates significant challenges for code understanding, embedded systems maintenance, and development team knowledge transfer.

\subsection{Problem Statement and Formal Definition}

\textbf{Definition 1 (Datasheet-to-Code Mapping Problem):} Given a datasheet document $S$ consisting of sections $S = \{s_1, s_2, ..., s_n\}$ and a code repository $R$ containing folders $F = \{f_1, f_2, ..., f_m\}$, files $Files = \{file_1, file_2, ..., file_k\}$, and code symbols $E = \{e_1, e_2, ..., e_l\}$, the datasheet-to-code mapping problem seeks to establish a function $M: S \rightarrow \mathcal{P}(E)$ that maps each datasheet section to a subset of relevant code symbols, where $\mathcal{P}(E)$ denotes the power set of $E$. \\ \\
\textbf{Definition 2 (Hierarchical Mapping):} A hierarchical mapping is a composition of functions $M = M_4 \circ M_3 \circ M_2 \circ M_1$ where:
\begin{itemize}
\item $M_1: S \rightarrow \mathcal{P}(F)$ maps datasheet sections to relevant folders
\item $M_2: S \times \mathcal{P}(F) \rightarrow \mathcal{P}(Files)$ maps sections and folders to relevant files  
\item $M_3: S \times \mathcal{P}(Files) \rightarrow \mathcal{P}(E)$ maps sections and files to code symbols
\item $M_4: S \times \mathcal{P}(E) \rightarrow \mathcal{P}(E) \times Status$ validates mapping and determines implementation status
\end{itemize}

\noindent \textbf{Definition 3 (Code Symbol Types):} Code symbols $e \in E$ are categorized into types $T = \{functions, \\
 macros, structs, constants, enums, typedefs\}$, where each element $e$ has an associated type $type(e) \in T$.

\medskip
\noindent The datasheet-to-code mapping problem manifests across multiple dimensions that existing approaches fail to address adequately:

\noindent\textbf{Scale and Hierarchical Relationships:} The mapping problem is inherently hierarchical, requiring analysis at multiple levels of granularity - from datasheet sections to repository folders, to implementation files, to specific code symbols. This hierarchical nature demands systematic decomposition rather than direct mapping approaches.

\noindent\textbf{Diverse Code Artifacts:} Embedded systems implementation involves various code artifacts beyond simple functions: configuration parameters and constants, data structure definitions (structs, typedefs, enums), preprocessor macros and compile-time definitions, protocol message formats and state machines, and hardware abstraction interfaces. Traditional function-focused approaches miss critical implementation components in systems-level programming.

\noindent\textbf{Dynamic Evolution:} Embedded systems repositories continuously evolve with new features, bug fixes, and refactoring. Manual mapping quickly becomes outdated, requiring automated approaches that can track not just what is implemented, but also what is missing or partially implemented relative to datasheet specifications.

\subsection{Limitations of Existing Approaches}

Current datasheet-to-code mapping approaches suffer from fundamental limitations:

\begin{itemize}
\item \textbf{Manual Processes:} Traditional approaches rely on expert analysis, which is time-consuming, error-prone, and does not scale to large embedded systems codebases
\item \textbf{Limited Scope:} Most tools focus on function-level mapping, overlooking macros, constants, and data structures critical to datasheet specification implementation
\item \textbf{Lack of Context:} Simple keyword matching fails to capture semantic relationships between datasheet requirements and code implementations
\item \textbf{No Gap Analysis:} Existing tools identify what exists but not what is missing relative to datasheet specifications
\end{itemize}









The remainder of this paper is organized as follows: Section~\ref{sec:related} reviews related work, Section~\ref{sec:baseline} presents the baseline approaches evaluated, Section~\ref{sec:methodology} describes the proposed methodology, Section~\ref{sec:evaluation} presents experimental results, Section~\ref{sec:applications} discusses applications, and Section~\ref{sec:conclusion} concludes with future directions.


\section{Related Work}
\label{sec:related}

The problem of establishing mapping between software specifications and implementations has been extensively studied in software engineering research, particularly in the area of Traceability Link Recovery (TLR). This section reviews existing approaches and positions the proposed methodology within the broader context of automated code understanding and specification-to-code mapping.

Traditional code understanding research has focused on establishing relationships between different software artifacts throughout the development lifecycle. Early foundational work identified the challenges of maintaining links between requirements and implementation artifacts, laying the groundwork for automated code mapping approaches. Subsequent work explored automated mapping recovery using information retrieval techniques, demonstrating the potential for automated approaches but highlighting limitations in precision and recall.

Information retrieval (IR) techniques have been widely applied to traceability link recovery and code mapping tasks. Vector Space Models (VSM) and Latent Semantic Indexing (LSI) have been used to establish semantic links between requirements and code. More recent work has explored the use of topic modeling techniques such as Latent Dirichlet Allocation (LDA) for code mapping. However, recent research by Wang et al.~\cite{wang2024traceability} has shown that traditional textual similarity approaches for TLR have reached their limits, demonstrating that purely text-based methods are insufficient for accurate traceability link recovery. Their comprehensive analysis reveals that LLM-based approaches can significantly outperform traditional textual similarity methods, provide empirical evidence supporting the use of LLMs.

Building on this insight, hybrid approaches combining information retrieval and machine learning have emerged as promising solutions. Guo et al.~\cite{guo2023information} proposed a hybrid method that integrates information retrieval techniques with machine learning for improved traceability link recovery, demonstrating better performance than purely IR-based approaches. However, these approaches typically focus on high-level document similarity rather than the fine-grained, hierarchical mapping required for precise specification-to-code understanding.

Static analysis tools have long been used for code understanding and analysis. Tools like commercial static analysis platforms, Doxygen~\cite{doxygen2024}, and various AST-based analyzers provide structural information about codebases. However, these tools typically lack the semantic understanding necessary to map natural language specifications to code implementations. Recent work has explored combining static analysis with natural language processing for improved code understanding.

The application of machine learning techniques to software engineering problems has gained significant attention. Deep learning models have been applied to code summarization, code generation, and bug detection. However, the specific problem of hierarchical specification-to-code mapping has received limited attention in the machine learning literature.

The evolution of neural language models has significantly advanced code understanding capabilities. Early transformer-based models like CodeBERT~\cite{feng2020codebert} and GraphCodeBERT~\cite{guo2021graphcodebert} established the foundation, while recent large language models~\cite{chen2021evaluating} have demonstrated even stronger performance on code understanding tasks. The findings by Wang et al.~\cite{wang2024traceability} show that LLM-based approaches significantly outperform traditional textual similarity methods for TLR provide strong empirical support for leveraging LLMs in specification-to-code mapping tasks. However, the application of LLMs to hierarchical specification-to-code mapping with comprehensive code coverage remains largely unexplored.

While existing research has advanced traceability link recovery and code understanding, key gaps remain. Most approaches attempt direct mapping without considering software hierarchy, focus only on functions rather than diverse code artifacts, and lack systematic gap identification. Our methodology addresses these limitations through a hierarchical approach that leverages LLMs for semantic understanding while maintaining comprehensive code coverage.

\section{Baseline Approaches and Comparative Analysis}
\label{sec:baseline}

This section presents the baseline approaches evaluated to establish performance benchmarks and identify limitations that motivated the development of the proposed hierarchical methodology. The evaluation process involved iterative improvements: starting with external grep-based tools, developing an improved BM25 + vector embeddings approach, and finally creating the hierarchical methodology presented in this work.

\subsection{Traditional Text-Based Search Methods}

Traditional text-based search methods represent the most straightforward approach to specification-to-code mapping, utilizing grep implementation (mgrep~\cite{mgrep2024}) to locate specification-relevant code through direct keyword matching. The technical implementation involves manual extraction of key terms from specification sections, followed by pattern matching across repository files using optimized grep variants. The system processes results by selecting top-K matches per query with file:line-range extraction and automated retrieval of matching code blocks from identified locations. This approach offers significant advantages in terms of fast execution speed, simple implementation requirements, and minimal computational overhead, making it particularly effective for direct keyword matching scenarios where obvious correspondences exist between specification terminology and code identifiers.

However, the fundamental limitations of purely syntactic matching without semantic understanding severely constrain the effectiveness of traditional text-based methods. The approach lacks hierarchical organization capabilities and fails to consider contextual relationships within software architecture, resulting in high false positive rates due to keyword ambiguity. When specification terms appear in multiple contexts throughout a codebase, grep-based methods cannot distinguish between functionally relevant matches and coincidental keyword occurrences, leading to mappings that are syntactically correct but semantically inappropriate for the intended specification requirements.

\subsection{BM25 + Vector Embeddings Approach}

To address the semantic limitations of traditional grep methods, an improved baseline approach was developed using a hybrid scoring system that combines BM25~\cite{robertson2009probabilistic} keyword matching with semantic vector embeddings~\cite{mikolov2013efficient} to map documentation chunks directly to code symbols. The technical architecture employs a sophisticated multi-stage pipeline beginning with PDF processing tools for converting specifications to clean markdown through HTML intermediation, followed by semantic chunking using sliding window embeddings with change-point detection to identify coherent document segments. Code analysis is performed through LSP server integration, which leverages language server protocol~\cite{microsoft2024lsp} capabilities to extract comprehensive symbol definitions including functions, classes, and types with their associated source code context. The system utilizes a dual-storage architecture combining LanceDB~\cite{lancedb2024recipes} for vector-based semantic search operations and Kuzu graph database~\cite{kuzu2024graph} for maintaining symbol relationship information, while BM25 indexes provide efficient keyword-based pre-filtering capabilities.

The approach employs a weighted combination of keyword relevance and semantic similarity scores that integrates multiple similarity signals through a computationally efficient search strategy. The semantic chunking process utilizes sliding window embeddings to create coherent document segments, applying change-point detection algorithms to identify natural boundaries in specification content. Symbol representation employs embedding models including e5-large-v2~\cite{wang2022e5large} (1024-dimensional) with L2 normalization for cosine similarity calculations. The hybrid scoring system combines BM25 keyword relevance with keyword overlap similarity and vector similarity using Jaccard coefficients. The search strategy implements a two-stage filtering approach: BM25 pre-filtering identifies the top 200 candidate symbols based on keyword relevance, followed by vector search refinement to select the final top 50 results, significantly reducing computational overhead while maintaining search quality.

This hybrid approach demonstrates significant improvements over traditional grep methods through its effective multi-modal analysis capabilities, scalable pre-filtering mechanisms, and enhanced semantic understanding of code-specification relationships. The integration of vector embeddings enables the system to capture semantic similarities that extend beyond exact keyword matches, while the BM25 component ensures that relevant terminology is appropriately weighted in the matching process. The two-stage filtering strategy provides computational efficiency by reducing the search space before expensive vector operations, while batch processing and caching mechanisms for BM25 indexes enable scalable deployment across large repositories. The flexible scoring system architecture allows for domain-specific weight adjustments, and the comprehensive metadata storage supports rich contextual analysis of matching results.

Despite these advances, the BM25 + embeddings approach has fundamental limitations. Vector averaging dilutes semantic coherence when multiple technical concepts are combined into single chunks, causing the averaged representation to lose precise meaning and produce topically related but functionally inappropriate matches. Meanwhile, BM25 keyword matching fails to capture semantic relationships and contextual nuances, making it unable to distinguish between similar terms used in different functional contexts.

The approach's primary weakness is its inability to leverage software repository structure, where folder organization and architectural layers provide crucial mapping context. While it improves semantic understanding over grep methods, it treats all code symbols as equivalent entities, ignoring their position within the architecture hierarchy. This becomes problematic in complex scenarios where component relationships are essential for identifying correct implementations at appropriate abstraction levels, resulting in matches that are semantically related but architecturally inappropriate.

\subsection{Example Results Comparison}

To illustrate the differences between baseline approaches and the hierarchical methodology, this section presents example results from the same specification query: ``Introduction / Initialization'' (Section 1) with query terms ``NCI interface DH NFCC NFCEE logical connection RF interface''. \\ \\
\textbf{Source Repository:} NXP NFC Linux Implementation~\cite{nxp2024libnfc} \\ \\
\textbf{Expected Output:} 
\begin{small}
\texttt{src/service, src/halimpl} $\rightarrow$ \texttt{service/nfc\_service.c, halimpl/phTmlNfc\_i2c.c} $\rightarrow$ \texttt{nfcService\_Init(), phTmlNfc\_I2COpen()}
\end{small}

\subsubsection{mgrep Results Analysis}

\textbf{Expected:} System initialization functions in service layer\\
\textbf{Found:} Runtime notification handling in HAL extension layer

The mgrep approach demonstrates fundamental limitations in distinguishing between functionally different code contexts despite keyword similarity. While the method was expected to identify initialization functions such as \texttt{nfcService\_Init()} within service layer components, it instead located runtime logic functions like \texttt{phNxpNciHal\_NfcDep\_rsp\_ext()} in HAL extension code. This represents a critical layer mismatch where service layer components were expected but Hardware Abstraction Layer extension code was found, along with a behavioral context mismatch between expected startup/initialization behavior and actual P2P priority logic and RF interface detection during runtime. The file location discrepancy further illustrates this issue, with the expected \texttt{service/nfc\_service.c} replaced by \texttt{halimpl/pn54x/hal/} \texttt{phNxpNciHal\_ext.c}. The root cause of these failures lies in keyword matching without semantic understanding of functional requirements, leading to identification of code containing relevant terms but serving entirely different purposes within the system architecture.

\subsubsection{BM25 + Vector Embeddings Results Analysis}

\textbf{Expected:} High-level service initialization functions\\
\textbf{Found:} Low-level I2C hardware configuration functions

The BM25 + vector embeddings approach, while demonstrating improved semantic understanding over grep methods, still exhibits significant abstraction level mismatches that highlight the limitations of direct chunk-to-symbol mapping strategies. The method was expected to identify service-level APIs such as \texttt{nfcService\_Init()} but instead located hardware-level configuration functions like \texttt{phTmlNfc\_i2c\_open\_} \texttt{and\_configure()}, representing a fundamental abstraction level disconnect. The functional scope diverged from expected comprehensive system initialization to specific I2C bus setup and pin configuration, while the component layer shifted from the intended service layer to Transport Mapping Layer (TML) components. The implementation focus similarly deviated from expected behavioral initialization logic to hardware interface setup and device reset sequences. This pattern of results demonstrates that while semantic understanding improved over grep methods, the direct chunk-to-symbol mapping approach still lacked sufficient hierarchical context to distinguish between different abstraction levels within the software architecture, leading to technically accurate but functionally inappropriate mapping.

\subsubsection{Hierarchical Methodology Results Analysis}

\textbf{Expected:} 
\begin{small}
\texttt{src/service, src/halimpl} $\rightarrow$ \texttt{service/nfc\_service.c, halimpl/phTmlNfc\_i2c.c} $\rightarrow$ \\
\texttt{nfcService\_Init(), phTmlNfc\_I2COpen()}
\end{small}\\
\textbf{Found:} 
\begin{small}
\texttt{src/service, src/halimpl} $\rightarrow$ \texttt{service/nfc\_service.c, halimpl/phTmlNfc\_i2c.c} $\rightarrow$ \\
\texttt{nfcService\_Init(), phTmlNfc\_I2COpen()}
\end{small}

The hierarchical methodology demonstrates strong accuracy across most levels of the specification-to-code mapping process, achieving close alignment between expected and found results with minor variations. The approach correctly identified both service and hardware implementation folders, demonstrating proper understanding of the dual-layer architecture required for NFC initialization. The file mapping precision located the primary files containing initialization logic at appropriate abstraction levels, specifically \texttt{service/nfc\_service.c} for high-level service coordination and \texttt{halimpl/phTmlNfc\_i2c.c} for hardware abstraction layer implementation, though some secondary initialization functions were found in related files within the same architectural layers. At the function level, the methodology successfully identified the core initialization functions \texttt{nfcService\_Init()} and \texttt{phTmlNfc\_I2COpen()}, along with some additional initialization-related functions that provide supporting functionality for the specified requirements. The hierarchical approach maintained understanding of system architecture from service layer to hardware abstraction, preserving the contextual relationships that enable accurate mapping across multiple architectural levels, with minor drift occurring in the selection of specific supporting functions within the correct architectural context.

\subsubsection{Comparative Analysis}

\textbf{Drift Score Definition:} The drift score measures how far the mapped results deviate from the expected structure across multiple levels: folder organization, file selection, and function type appropriateness. A score of 0.0 indicates perfect alignment with expected results, while higher scores indicate greater structural deviation from the target mapping.

The drift score quantifies the distance between expected and actual mapping across folder structure, file selection, and functional appropriateness.

\begin{table}[ht]
\centering
\begin{tabular}{lcccc}
\toprule
\textbf{Method} & \textbf{Layer Match} & \textbf{Function Type} & \textbf{Drift Score} & \textbf{Expected Match} \\
\midrule
mgrep & HAL Extension & Runtime Logic & 2.5 & \texttimes \\
BM25 + Embeddings & Transport Layer & Hardware Config & 1.5 & \texttimes \\
Hierarchical & Service + HAL & Initialization & 0.3 & \checkmark \\
\bottomrule
\end{tabular}
\caption{Code-level comparison of baseline method results with mapping drift scores}
\end{table}

\textbf{Key Insights:} Baseline methods consistently found semantically related code but at wrong abstraction levels or functional contexts, resulting in significant mapping drift from expected results. The mgrep approach found keyword matches in runtime logic rather than initialization code, achieving a drift score of 2.5 due to incorrect file location, function type, and architectural layer mismatches. The BM25 approach improved semantic understanding but focused on hardware configuration rather than service initialization, resulting in a drift score of 1.5 with better architectural proximity but still incorrect functional context. The hierarchical approach successfully navigated from specification intent to implementation through systematic decomposition, achieving minimal drift (0.3) with minor variations in specific function selection while maintaining proper context across software architecture layers. The drift metric provides a measurable assessment of mapping accuracy beyond simple binary success/failure evaluation.

\section{Proposed Methodology}
\label{sec:methodology}

This section presents the hierarchical mapping methodology that systematically decomposes the specification-to-code mapping problem into manageable, sequential phases. Each step operates at a different level of granularity, enabling precise and scalable analysis while leveraging the natural organization of software repositories.
The methodology is implemented as a deterministic, multi-stage analysis pipeline, where each stage restricts the search space for subsequent stages.
\subsection{Methodology Overview}

The proposed approach addresses the limitations identified in existing methods through a hierarchical decomposition strategy. The methodology consists of four sequential steps:

\begin{enumerate}
\item \textbf{Folder Discovery:} Mapping specification sections to relevant repository folders
\item \textbf{File Discovery:} Identifying specific files within relevant folders  
\item \textbf{Code Symbol Discovery:} Extracting specific code symbols from relevant files
\item \textbf{Validation \& Gap Analysis:} Validating mapping and identifying implementation gaps
\end{enumerate}

\begin{algorithm}
\caption{Hierarchical Specification-to-Code Mapping}
\begin{algorithmic}[1]
\REQUIRE Specification $S = \{s_1, s_2, ..., s_n\}$, Repository $R$
\ENSURE Mapping $M: S \rightarrow \mathcal{P}(E)$, Status assessments
\STATE Generate repository structure documentation $D_R$
\FOR{each section $s_i \in S$ (parallel execution)}
    \STATE $F_i \leftarrow \text{FolderDiscovery}(s_i, D_R)$
    \STATE $Files_i \leftarrow \text{FileDiscovery}(s_i, F_i)$
    \STATE $E_i \leftarrow \text{CodeElementDiscovery}(s_i, Files_i)$
\ENDFOR
\FOR{each section $s_i \in S$ (sequential execution)}
    \STATE $(E'_i, status_i) \leftarrow \text{Validate}(s_i, E_i, context_{i-1})$
    \STATE $M(s_i) \leftarrow E'_i$
\ENDFOR
\RETURN $M$, $\{status_1, status_2, ..., status_n\}$
\end{algorithmic}
\end{algorithm}

This decomposition enables the system to leverage repository structure and organization for improved accuracy while maintaining comprehensive coverage of all code artifacts.

\subsection{Step 1: Folder Discovery}

\textbf{Objective:} Identify which folders in the repository are likely to contain implementations relevant to each specification section. \\ \\
\textbf{Formal Definition:} Given a specification section $s_i$ and repository structure documentation $D_R$, the folder discovery function is defined as:
\begin{equation}
M_1(s_i, D_R) = \{f_j \in F : similarity(s_i, desc(f_j)) > \theta_1\}
\end{equation}
where $desc(f_j)$ represents the generated description of folder $f_j$ and $\theta_1$ is the similarity threshold. \\ \\
\textbf{Key Techniques:}
\begin{itemize}
\item \textbf{Adaptive Chunking:} Dynamic processing of large repositories with parallel chunk analysis
\item \textbf{Parent-Child Filtering:} Automatic removal of redundant folder relationships, preferring most specific folders
\item \textbf{Refinement Selection:} Final filtering step to select most relevant folders from multiple candidates
\end{itemize}

\subsection{Step 2: File Discovery}

\textbf{Objective:} Within identified folders, determine which specific files contain relevant implementations. \\ \\
\textbf{Formal Definition:} Given a specification section $s_i$ and relevant folders $F_i = M_1(s_i, D_R)$, the file discovery function is:
\begin{equation}
M_2(s_i, F_i) = \{file_k \in Files : file_k \in folder(F_i) \land relevance(s_i, file_k) > \theta_2\}
\end{equation}
where $folder(F_i)$ returns all files within folders $F_i$ and $\theta_2$ is the relevance threshold. \\ \\
\textbf{Implementation:} The system employs on-demand structure generation to create comprehensive file documentation. For each relevant folder, LLM-based analysis using models such as Gemini 2.5 Flash~\cite{gemini2024team} generates \texttt{folder\_structure.md} files containing detailed descriptions of each file's purpose and functionality. A caching mechanism avoids regenerating existing structure files, while thread-safe processing handles concurrent folder analysis efficiently.

\subsection{Step 3: Code Symbol Discovery}

\textbf{Objective:} Within relevant files, identify specific code symbols (functions, macros, structs, constants) that implement specification requirements. \\ \\
\textbf{Formal Definition:} Given a specification section $s_i$ and relevant files $Files_i = M_2(s_i, F_i)$, the code symbol discovery function is:
\begin{equation}
M_3(s_i, Files_i) = \{e_j \in E : e_j \in elements(Files_i) \land semantic\_match(s_i, e_j) > \theta_3\}
\end{equation}
where $elements(Files_i)$ extracts all code symbols from files $Files_i$ using static analysis, and \\  $semantic\_match$ evaluates semantic relevance. \\ \\
\textbf{Implementation:} The system integrates Universal Ctags~\cite{universalctags2024} for reliable parsing of C/C++ code symbols, extracting functions, macros, structs, constants, and enums with precise line number information. The current implementation focuses on C/C++ codebases, leveraging language-specific parsing capabilities for optimal accuracy. Automated code summarization generates compact representations preserving key elements while significantly reducing token usage. Thread-safe generation creates \texttt{\{filename\}\_\{ext\}\_structure.md} files with caching to avoid redundant processing of previously analyzed files.

\subsection{Step 4: Validation \& Gap Analysis}

\textbf{Objective:} Validate the complete mapping, determine implementation status, and identify gaps. \\ \\ 
\textbf{Formal Definition:} The validation function processes mapping sequentially to maintain context:
\begin{equation}
M_4(s_i, E_i, context_{i-1}) = (E'_i, status_i)
\end{equation}
where $E'_i \subseteq E_i$ represents the refined set of code symbols and $status_i$ represents the implementation status:
\begin{align}
status_i \in \{&Implemented, Partially\_Implemented, \nonumber\\
&Not\_Implemented, Not\_Applicable\}
\end{align}

%
%
%

\begin{figure}[ht]
\centering
\includegraphics[width=0.95\linewidth]{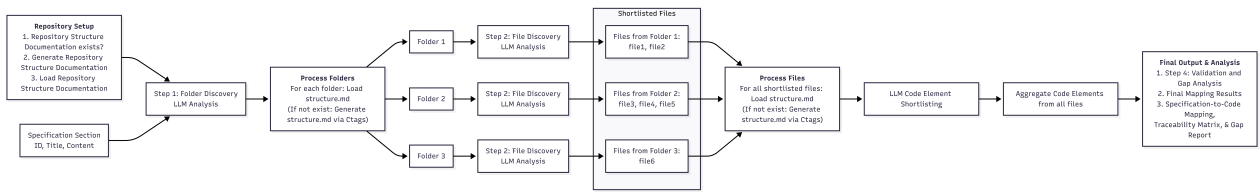}
\caption{Flow diagram illustrating the hierarchical mapping process for a single specification section, showing the systematic progression from specification analysis through folder discovery, file identification, and code symbol extraction to final validation.}
\label{fig:methodology_flow}
\end{figure}

\section{Experimental Evaluation}
\label{sec:evaluation}

This section presents a comprehensive evaluation of the proposed methodology through comparison with baseline approaches. The evaluation demonstrates the progressive improvements achieved through iterative development and provides empirical evidence for the design decisions made throughout the development process.

\subsection{Experimental Setup and Evaluation Metrics}

\textbf{Dataset:} The evaluation was conducted on 154-156 specification sections across multiple embedded systems protocol repositories. The dataset was selected to represent diverse specification types and implementation patterns commonly found in embedded systems development.

\textbf{Evaluation Metrics:} Six primary metrics were used to assess methodology performance:
\begin{itemize}
\item \textbf{Confidence Score:} Average LLM-assigned confidence for mapped elements (0-1 scale)
\item \textbf{Code Symbol Coverage:} Average number of code symbols mapped per specification section
\item \textbf{Runtime Performance:} Total processing time in minutes for all specification sections
\item \textbf{Token Consumption:} Total number of LLM tokens consumed during mapping
\item \textbf{File Existence Accuracy:} Percentage of mapped files that actually exist in the repository
\item \textbf{File Mapping Accuracy:} Percentage of desired files correctly mapped for each section
\end{itemize}

\subsection{Experimental Results}

The experimental evaluation demonstrates significant improvements across all metrics. Table~\ref{tab:results} summarizes the performance comparison across seven approaches including baseline methods, while the detailed performance trends are shown across all evaluation metrics in the accompanying figures.

\begin{table}[!t]
\caption{Comprehensive Performance Comparison}
\centering
\footnotesize
\setlength{\tabcolsep}{2pt}
\begin{tabular}{lcccccc}
\toprule
\textbf{Method} & \textbf{Confidence} & \textbf{Elements} & \textbf{Runtime} & \textbf{Tokens} & \textbf{File Exist.} & \textbf{File Map.} \\
& \textbf{(\%)} & \textbf{per Section} & \textbf{(min)} & \textbf{(M)} & \textbf{(\%)} & \textbf{Acc. (\%)} \\
\midrule
mgrep & N/A & N/A & N/A & N/A & 100.0 & 0.0 \\
BM25 & N/A & N/A & N/A & N/A & 100.0 & 0.0 \\
Gemini$^{*}$ & 89.5 & 19.0 & 90 & 68.8 & 92.7 & 53.3 \\
Gemini$^{*}$ + H2LooP Toolchain (Structures) & 90.4 & 15.9 & 21 & 49.0 & 93.3 & 63.3 \\
Gemini$^{*}$ + H2LooP Toolchain (Structures + Ctags) & 90.1 & 16.4 & 24 & 17.7 & 95.3 & 56.7 \\
Qwen3$^{**}$ + H2LooP Toolchain (Structures + Ctags) & 84.1 & 11.6 & 10 & 3.2 & 96.3 & 66.7 \\
\textbf{Proposed (Full)} & \textbf{83.1} & \textbf{9.1} & \textbf{18} & \textbf{10.9} & \textbf{95.9} & \textbf{73.3} \\
\bottomrule
\end{tabular}
\begin{flushleft}
\small
$^{*}$Gemini 2.5 Flash \\
$^{**}$Qwen3-Coder-30B-A3B-Instruct-FP8
\end{flushleft}
\label{tab:results}
\end{table}
\FloatBarrier

\begin{figure}[!t]
\centering
\includegraphics[width=0.7\textwidth]{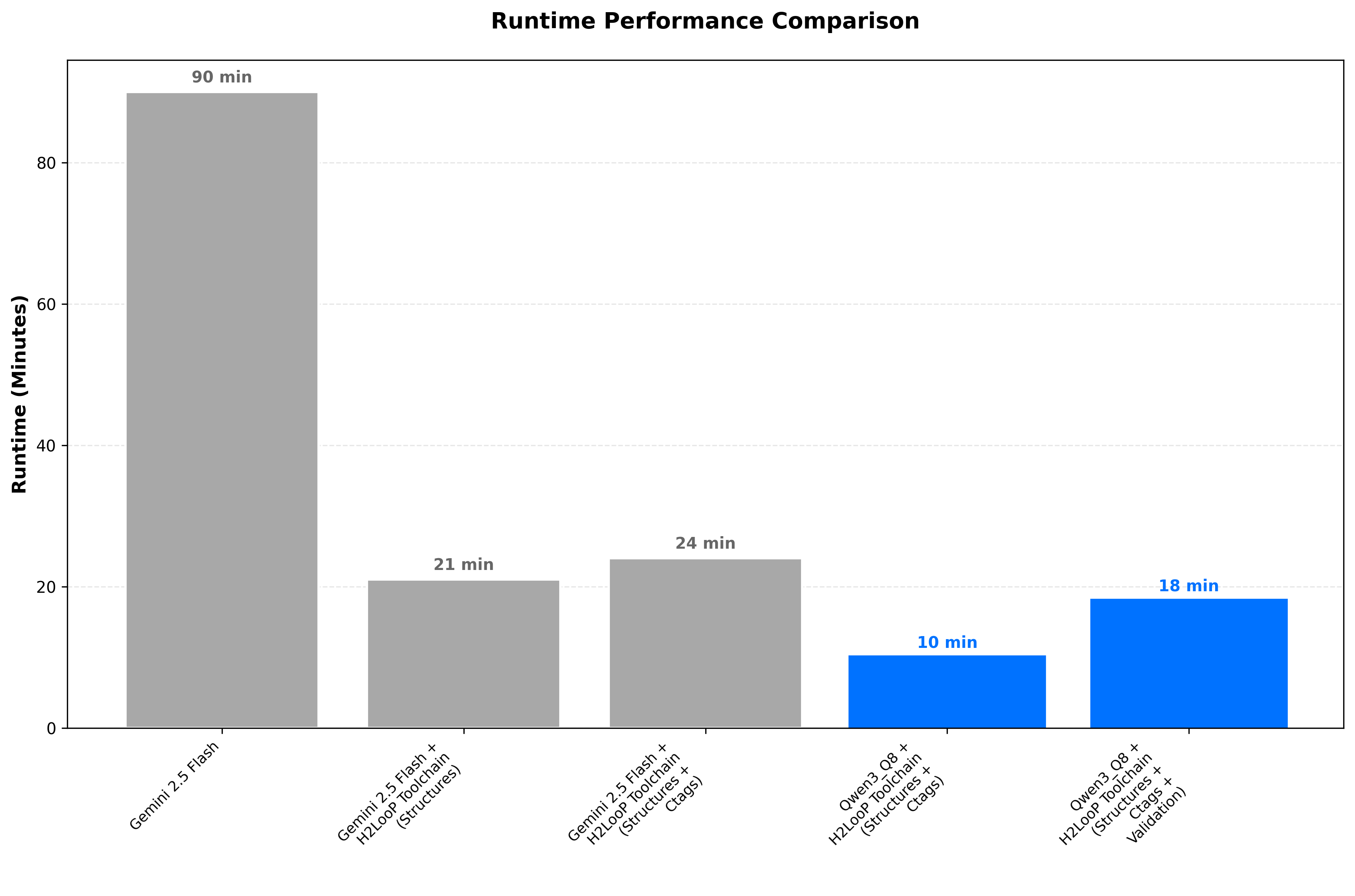}
\caption{Runtime performance comparison showing processing efficiency improvements through iterative methodology development.}
\label{fig:runtime}
\end{figure}

\begin{figure}[!t]
\centering
\includegraphics[width=0.7\textwidth]{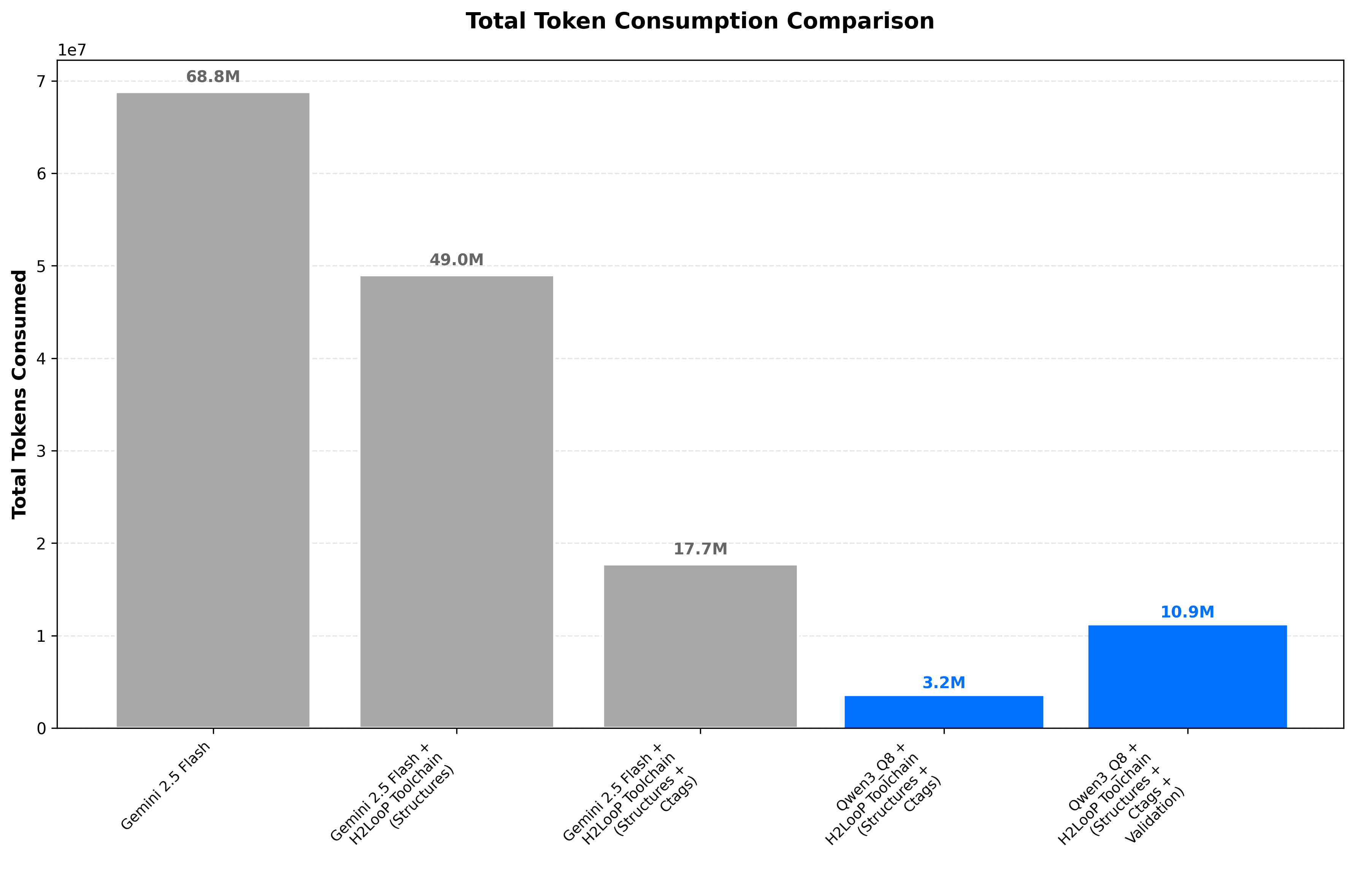}
\caption{Token consumption comparison illustrating computational cost optimization achieved by the proposed methodology.}
\label{fig:tokens}
\end{figure}

\begin{figure}[!t]
\centering
\includegraphics[width=0.7\textwidth]{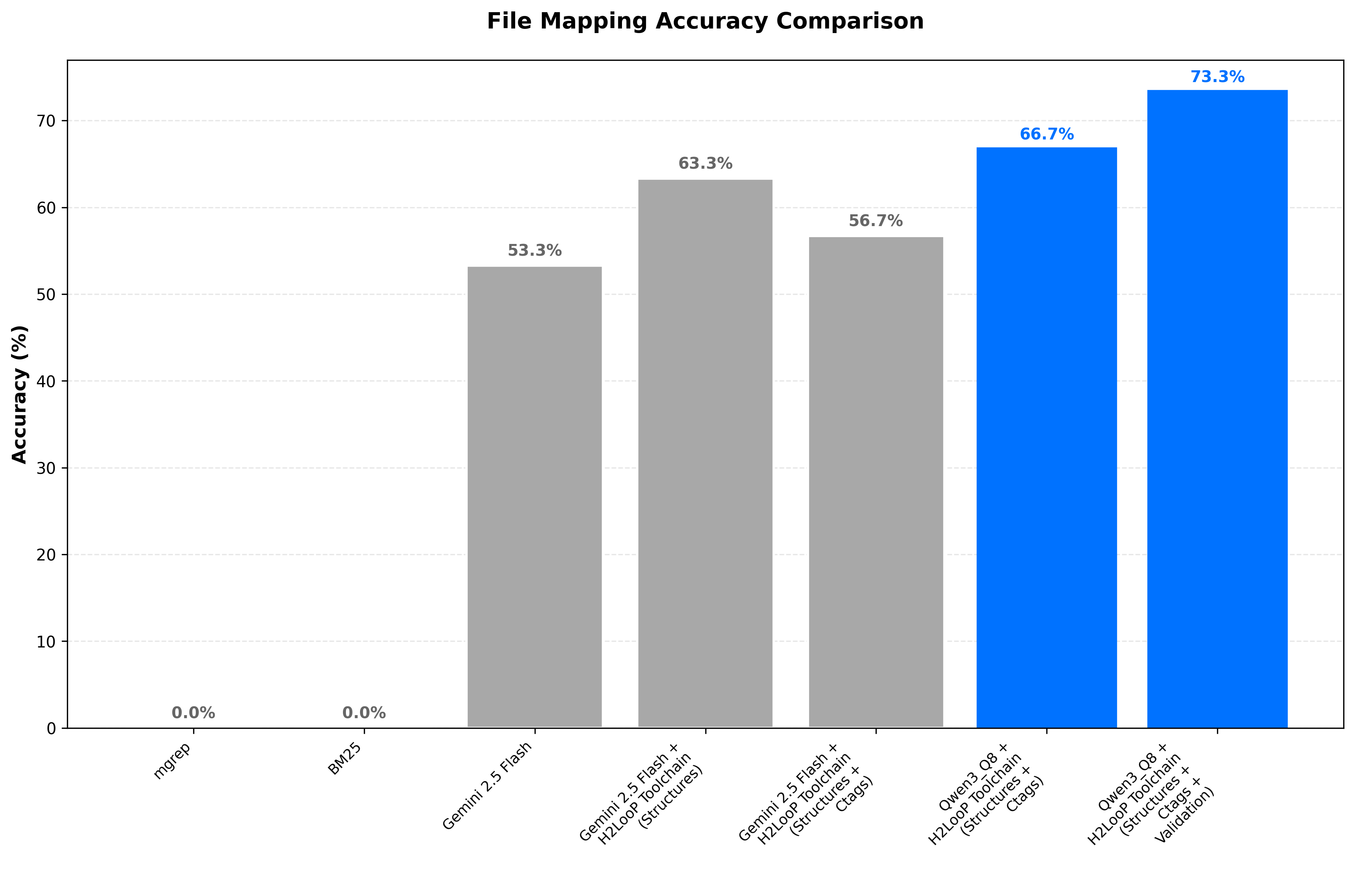}
\caption{File mapping accuracy comparison showing the percentage of desired files that were correctly mapped. Statistical approaches (mgrep, BM25) achieve perfect file existence but zero mapping accuracy, while the hierarchical approach demonstrates progressive improvement.}
\label{fig:file_mapping}
\end{figure}

\subsection{Discussion of Results}

\textbf{Statistical vs. Semantic Approaches:} The baseline statistical approaches (mgrep, BM25) demonstrate perfect file existence accuracy (100\%) as they only reference files that actually exist in the repository. However, they achieve zero file mapping accuracy (0\%) because they map semantically related but functionally incorrect files. This highlights the fundamental limitation of keyword-based and statistical similarity approaches.

\textbf{Structural Context and Code Parsing:} The addition of repository, folder and file level structures (Iteration 2) provided the most significant runtime improvement (76.7\% reduction) by giving LLMs essential context for accurate folder and file identification, eliminating exploratory analysis. This improved file mapping accuracy from 53.3\% to 63.3\%, while pre-validation confidence remained elevated at 90.4\%. Ctags integration (Iteration 3) resulted in the largest reduction in token consumption (63.8\% reduction) through compact code representation, while maintaining mapping accuracy through reliable code symbol extraction. However, pre-validation confidence remained high at 90.1\%, indicating potential calibration issues.

\textbf{Model Selection Optimization:} The transition to Qwen3-Coder-30B-A3B-Instruct-FP8 (Iteration 4) achieved the lowest token consumption among evaluated configurations with 81.9\% token reduction and 58.3\% runtime improvement, demonstrating the value of domain-specific, self-hosted models for code analysis tasks. This significant token reduction is partly attributed to switching from Gemini 2.5 Flash, which is a thinking model that involves internal reasoning processes, to Qwen3. File mapping accuracy improved to 66.7\% while maintaining high file existence accuracy (96.3\%). This iteration showed improved confidence calibration with a pre-validation confidence of 84.1\%, closer to realistic levels compared to earlier Gemini-based approaches.

\textbf{Validation Quality Enhancement:} Sequential validation (Iteration 5) introduces controlled overhead (80\% runtime increase from Iteration 4) but provides essential gap analysis and implementation status assessment, transforming the system from discovery to comprehensive analysis. This achieves the highest file mapping accuracy (73.3\%) while maintaining excellent file existence accuracy (95.9\%). The validation step improves confidence calibration, with the final methodology achieving a more realistic confidence score of 83.1\% compared to pre-validation confidence levels that exhibited inflated estimates (89.5-90.4\% for Gemini-based approaches, 84.1\% for Qwen3).

\subsection{Performance Summary}

The proposed methodology demonstrates substantial improvements across all evaluation metrics, establishing its effectiveness for practical specification-to-code mapping applications. The methodology achieves an average confidence score of 83.1\% through rigorous validation processes, indicating high reliability in generated mappings while maintaining comprehensive coverage with an average of 9.1 precisely mapped code symbols per specification section.

The efficiency improvements are particularly noteworthy, with the methodology achieving an 84\% reduction in computational overhead, decreasing token consumption from 68.8 million to 10.9 million tokens compared to baseline approaches. Simultaneously, processing time is reduced by 80\%, from 90 minutes to 18 minutes, making the approach viable for large-scale industrial applications. These efficiency gains are achieved without compromising accuracy, as evidenced by the 95.9\% file existence accuracy and 73.3\% file mapping accuracy that significantly outperforms statistical approaches which achieve 0\% mapping accuracy despite perfect file existence scores.

\section{Applications and Impact}
\label{sec:applications}

The proposed methodology addresses specification-to-code traceability challenges across multiple domains in software development and AI research. The systematic mapping between specification requirements and code implementations creates high-quality, structured datasets valuable for training code understanding and generation models. These training pairs capture semantic relationships between natural language descriptions and code artifacts, enabling more accurate AI systems for automated code analysis, documentation generation, and specification-driven development. With 83.1\% confidence while maintaining broad coverage, the approach provides a solid foundation for constructing training datasets that advance code intelligence systems.

Organizations developing standards-compliant software can leverage this approach for systematic verification of specification coverage and regulatory compliance. The gap analysis component identifies missing implementations, partial coverage areas, and compliance gaps with precise granularity, supporting critical certification processes in automotive, aerospace, and medical device industries where specification adherence is mandatory. The methodology enables auditors and compliance teams to trace requirements from high-level specifications down to specific code symbols, providing clear evidence of implementation completeness.

The mapping capability provides a systematic method for understanding large, complex codebases through their specification requirements, addressing significant challenges in software maintenance and evolution. This proves particularly valuable for legacy system analysis where original documentation may be incomplete, enabling development teams to reconstruct relationships between intended functionality and actual implementation. The approach supports knowledge transfer by creating comprehensive mappings that help new team members understand system architecture and implementation patterns, while enhancing code review processes by providing specification context for changes.

\section{Conclusion and Future Work}
\label{sec:conclusion}

This work advances automated specification-to-code mapping through a decomposition approach that systematically maps from specifications to folders, files, and code symbols. The proposed methodology demonstrates significant improvements in both accuracy and efficiency over existing approaches, achieving high confidence while reducing computational overhead and processing time. Key innovations include multi-level decomposition for precise analysis and comprehensive code coverage extending beyond functions to include macros, structs, constants, and configuration parameters.

Experimental evaluation across multiple embedded systems repositories shows substantial improvements in mapping accuracy and computational efficiency compared to baseline methods. The methodology processes specification sections with precisely mapped code symbols while maintaining excellent file existence accuracy, providing a practical solution for automated specification-to-code understanding. The integration of LLM-based semantic analysis with automated structure generation creates a robust framework that respects software architecture principles while delivering precise results.

Future research directions include integration with development workflows for real-time compliance monitoring within continuous integration pipelines, enabling automated specification adherence verification during development. The generated mapping datasets provide valuable training data for specialized machine learning models, potentially leading to more sophisticated understanding systems that combine structural decomposition benefits with learned patterns from large-scale mapping examples.

The methodology represents a significant advancement in automated software analysis, providing a scalable and practical solution for establishing traceability between specifications and implementations. The approach demonstrates that systematic decomposition combined with LLM-based semantic understanding can overcome traditional text-based mapping limitations, enabling effective analysis of specification-implementation relationships in complex software systems.

\bibliographystyle{unsrt}  
\bibliography{references}

\end{document}